\documentclass{article}%
\usepackage{graphicx}
\usepackage{amsmath}%
\usepackage{amsfonts}%
\usepackage{amssymb}
\newtheorem{theorem}{Theorem}
\newtheorem{acknowledgement}[theorem]{Acknowledgement}

\begin{document}

\title{The Arrow of Time in Rigged Hilbert Space Quantum Mechanics}
\author{Robert C. Bishop$^{a,b}$\\$^{a}$Abteilung f\"{u}r Theorie und Datenanalyse, Institut\\f\"{u}r Grenzgebiete der Psychologie, Wilhelmstrasse\\3a, D-79098 Freiburg, Germany\\$^{b}$Department of Philosophy,\\Logic and Scientific Method, The London School\\of Economics, Houghton St., London,\\WC2A 2AE, United Kingdom}
\maketitle

\begin{abstract}
Arno Bohm and Ilya Prigogine's Brussels-Austin Group have been working on the
quantum mechanical arrow of time and irreversibility in rigged Hilbert space
quantum mechanics. A crucial notion in Bohm's approach is the so-called
preparation/registration arrow. An analysis of this arrow and its role in
Bohm's theory of scattering is given. Similarly, the Brussels-Austin Group
uses an excitation/de-excitation arrow for ordering events, which is also
analyzed. The relationship between the two approaches is initially discussed
focusing on their semi-group operators and time arrows. Finally a possible
realist interpretation of the rigged Hilbert space formulation of quantum
mechanics is considered.

\begin{acknowledgement}
This paper has benefitted from numerous discussions with A. Bohm, H.
Atmanspacher, I. Antoniou, F. Kronz and F. Schroeck.

\end{acknowledgement}
\end{abstract}

\noindent\pagebreak 

\section{Introduction}

When Dirac introduced his formalism for quantum mechanics (1981/1930), it
lacked a rigorous mathematical foundation. Von Neumann's pioneering work on
Hilbert space theory (1955/1932) became the mathematical foundation for
quantum mechanics (QM). Nevertheless, many physicists preferred using Dirac's
bra-ket formalism because of its calculational convenience among many other
advantages including: (1) observables can be treated as continuous operators,
(2) Hermitian observables have a complete set of eigenkets and their
corresponding eigenvalues can be discrete or continuous, and (3) state vectors
are well-behaved smooth functions. However, a rigorous justification for
Dirac's formalism cannot be given within Hilbert space (HS).

There are additional reasons to extend the HS formulation of QM to a broader
mathematical framework such as a rigged Hilbert space (RHS), also known as a
Gel'fand triplet or equipped space (Gel'fand and Vilenkin 1964; Gel'fand and
Shilov 1967; Bohm 1967; Bohm and Gadella 1989; Nagel 1989). After briefly
reviewing RHS (\S2), some reasons for going beyond HS will be given,
particularly scattering and decay phenomena (\S\S3-6). Finally, some initial
thoughts toward a realist interpretation of RHS QM are considered (\S\S7-8).

\section{What Is RHS?}

Let $\Psi$ be an abstract linear scalar product space and complete $\Psi$ with
respect to two topologies. The first topology is the standard HS topology
$\tau_{\mathcal{H}}$ defined by the norm%
\begin{equation}
\Vert h\Vert=\sqrt{(h,h)}%
\end{equation}
where $h$ is an element of $\Psi$. The second topology $\tau_{\Phi}$ is
defined by a countable set of norms%
\begin{equation}
\Vert\phi\Vert_{n}=\sqrt{(\phi,\phi)_{n}},\;n=0,1,2,...
\end{equation}
where $\phi$ is also an element of $\Psi$ and the scalar product in (2) is
given by
\begin{equation}
(\phi,\phi^{\prime})_{n}=(\phi,(\Delta+1)^{n}\phi^{\prime}),\;n=0,1,2,...
\end{equation}
where $\Delta$ is the Nelson operator $\Delta=$ $\sum_{i}\chi_{i}^{2}$. The
$\chi_{i}$ are the generators of an enveloping algebra of observables for the
system in question and they form a basis for a Lie algebra (Nelson 1959; Bohm
\textit{et al}. 1999). For example if we are modeling the harmonic oscillator,
the $\chi_{i}$ would be the position and momentum operators or, alternatively,
the raising and lowering operators (Bohm 1978). Furthermore if the operator
$\Delta+1$ is \textit{nuclear }then the space $\Phi$ defined by (2) is a
nuclear space (Bohm 1967; Treves 1967).

We obtain a Gel'fand triplet by completing $\Psi$ with respect to $\tau_{\Phi
}$ to obtain $\Phi$ and with respect to $\tau_{\mathcal{H}}$ to obtain
$\mathcal{H}$. In addition we consider the dual spaces of continuous linear
functionals $\Phi^{\times}$ and $\mathcal{H}^{\times}$ respectively. Since
$\mathcal{H}$ is self dual, we obtain%
\begin{equation}
\Phi\subset\mathcal{H}\subset\Phi^{\times}\text{ .}%
\end{equation}

The Nelson operator fully determines the space $\Phi$. However, there are many
inequivalent irreducible representations of an enveloping algebra of a group
characterizing a physical system (Bohm \textit{et al}. 1999). Therefore
further restrictions may be required to obtain a realization for $\Phi$. The
particular characteristics of the physical context of the system being modeled
provide some restrictions analogous to the situation for GNS representations
for the construction of $W^{\ast}$-algebras in algebraic quantum mechanics.
Additional restrictions may be required due to the convergence properties
desired for test functions in $\Phi$. In general one chooses the weakest
topology such that the algebra of operators for the physical problem is
continuous and $\Phi$ is nuclear. The physical symmetries of the system play
an important role in such choices (Bohm \textit{et al}. 1999).\footnote{In the
simple example of the harmonic oscillator, choosing the raising and lowering
operators as the generators for the algebra or the position and momentum
operators as the generators would yield different Nelson operators, but the
results are physically equivalent. However in general one does not get
physically equivalent results (e.g. choosing a different value of $j$ in the
rotation group corresponds to a different physical system/situation). So one
has to look at the symmetries, boundary conditions, causal mechanisms,
\textit{etc}. in order to decide which representation of an enveloping algebra
to use as a representation.}

In many regards, working in a RHS is only as complicated as using standard
Dirac bra-ket notation. There is an additional conceptual problem introduced
in the RHS extension to QM that is not present in the ordinary HS formulation;
namely, the choice of riggings $\Phi$ and $\Phi^{\times}$ are problem
dependent. Every physical system, or at best classes of systems as in
scattering, has its own RHS distinguished by the algebra of observables. This
problem does not exist in HS where the natural norm topology is prescribed for
all physical systems.

The typical choice for a realization of HS is the space of equivalence classes
of Lebesgue square integrable functions $L^{2}$. Smooth functions are defined
for every point, but the equivalence classes of $L^{2}$ functions, the vectors
of the HS, are not so defined. In RHS there are no equivalence class problems.
The vectors of $\Phi$ are functions that are defined point-wise and are
typically Riemann integrable. While the position and momentum operators do not
have eigenvectors in HS (Gel'fand and Shilov 1967), all eigenstates are
well-defined in RHS.\footnote{Hence, in RHS the observables form an algebra on
the entire space of physical states (including $\Phi^{\times}$, where Dirac
kets reside).}

More generally, RHS contains observables with continuous or even complex
eigenvalues, whereas HS does not, because the dual space $\Phi^{\times}%
$\ contains the appropriate eigenvectors along with distributions. This means
that the basis vector expansion of eigenvectors (Dirac's spectral
decomposition) can be given a rigorous foundation resulting in the nuclear
spectral theorem:
\begin{equation}
|\phi\rangle=\sum_{n}|E_{n})(E_{n}|\varphi)+\int|E\rangle\langle
E|\varphi\rangle d\mu(E)\text{.}%
\end{equation}
Here the rounded bras and kets denote elements definable on HS and the first
term in (5) represents the discrete part of the spectrum. The angular bras and
kets $\langle E|$, $|E\rangle$ denote elements defined in $\Phi^{\times}$,\ so
the second term in (5) represents the continuous part of the spectrum.

\section{Extrinsic vs. Intrinsic Irreversibility}

Solutions of Schr\"{o}dinger's equation in HS describe the temporal evolution
of isolated quantum systems in a time-reversible manner, yet many quantum
systems exhibit irreversible behavior (e.g. resonance and decaying states).
There are two distinct ways of describing irreversible processes (e.g.
Atmanspacher \textit{et al.} 2002). Irreversible behavior in quantum systems
is usually viewed as solely due to the interaction of a system with its
environment. This approach to irreversibility is described as
\textit{extrinsic}, because the environment is crucial for irreversible
evolution. Examples of extrinsic irreversibility are given by any open-system
described by a master equation. By contrast \textit{intrinsic} irreversibility
refers to irreversible behavior generated by the dynamics of a system; that is
to say, the Hamiltonian describes this irreversibility without explicit
reference to an environment. An example of intrinsic irreversibility is kaon decay.

Intrinsic irreversibility is of prime interest to Bohm and his collaborators,
as well as to Prigogine's Brussels-Austin Group, because these types of
irreversible processes are related to arrows of time. HS QM cannot give a
rigorous description of these types of physical processes. One reason is that
no HS elements exist whose survival probability has the right form of
exponential decay:
\begin{equation}
P_{s}=|(\phi,e^{-iHt}\phi)|^{2}\varpropto e^{-\Gamma t}\;.
\end{equation}
It might be objected that physical systems decay with deviations from
exponential decay which are too small to be measured experimentally. After all
probabilities like $P_{s}$ are observed in the laboratory as ratios of large
numbers $N(t)/N(0)$, where $N(t)$ is the number of counts of a detector and
$N(0)$ is the total population under observation. Unless the predicted
deviations from (6) are of time scales comparable to $1/\Gamma$, they cannot
be empirically observed. Recently deviations from exponential decay over short
time scales for atoms undergoing quantum tunneling have been reported
(Wilkinson \textit{et al.}, 1998).

When decaying states are involved, (5) can be rewritten as%
\begin{equation}
|\phi\rangle=\sum_{n}^{N}|\psi_{n}^{G}\rangle\langle\psi_{n}^{G}%
|\varphi\rangle+\int|E\rangle\langle E|\varphi\rangle d\mu(E)
\end{equation}
where $|\phi\rangle$ represents the prepared state vector and $|\psi_{n}%
^{G}\rangle$, the so-called Gamow vector\footnote{As originally introduced,
Gamow vectors were problematic in HS: their position probability density
increased exponentially for large negative values of $t$. But since decay
processes must begin at some past time $t=0$, a RHS removes this physically
problematic feature by allowing for more realistic boundary conditions (Bohm
\textit{et al}. 1997).}, represents decaying states. The first term on the rhs
of (7) represents a subdomain of the decaying state (note these are not
elements of $\mathcal{H}$, hence no rounded brakets). There is usually only a
small number $N$ at the available scattering energies. The second term
represents the background integral. The standard Weisskopf-Wigner
approximation amounts to ignoring the background integral, but the preparation
process does not always make this term negligible. The background integral
\textit{does not} have exponential time behavior, so if this term is
substantial, deviations from exponential decay will result. An effect of the
background amplitude is often observed in resonance scattering experiments,
whereas for decaying states, it is often neglected (Bohm 1994). So Wilkinson
\textit{et al}.'s observations can be explained as the effect of the
background integral in (7). Their experiment involved a series of
interventions (preparations) in the form of varying electromagnetic potentials
to introduce variations in the acceleration of the atoms under observation
leading to an \textit{extrinsically} irreversible decay process.

Another reason for which intrinsic irreversibility cannot naturally be
described in HS is that HS evolution is given by
\begin{equation}
U(t)|\phi(0)\rangle=e^{-iHt}|\phi(0)\rangle
\end{equation}
where $U(t)$ is a unitary group generated by the Hamiltonian $H$ for the
system. The operator $U(t)$ is a continuous operator with respect to the
topology $\tau_{\mathcal{H}}$ and forms a one-parameter group of operators.
The inverse is defined as $U^{-1}(t)=U(-t)$ for all $-\infty<t<\infty$, so the
evolution governed by $U(t)$ is time symmetric. However, semigroup operators
lack an inverse. Therefore semigroups of operators are the appropriate
operators for the evolution of intrinsically irreversible processes. In HS we
must appeal to interactions with an environment (i.e. extrinsic
irreversibility), whereas in RHS semigroup evolution and intrinsic
irreversibility naturally arise.

If $U(t)$ is a unitary operator on $\mathcal{H}$ and $\Phi\subset
\mathcal{H}\subset\Phi^{\times}$, then $U$ can be \textit{extended} to
$\Phi^{\times}$\ provided that (1) $U$ leaves $\Phi$ invariant, i.e. $U$:
$\Phi\rightarrow\Phi$, and (2) $U$ is continuous on $\Phi$ with respect to the
topology $\tau_{\Phi}$. The operator $U^{\times}$\ denotes the
\textit{extension} of the HS operator $U$ to $\Phi^{\times}$\ and is defined
by $\langle U\phi|F\rangle=\langle\phi|U^{\times}F\rangle$ for all $\phi
\in\Phi$ and $F\in\Phi^{\times}$. Additionally $U^{o}$: $\Phi\rightarrow\Phi$
denotes the \textit{restriction} of the HS operator $U$ to $\Phi$.

\section{Scattering}

Bohm and his co-workers have studied simple scattering experiments using RHS
(e.g. Bohm \textit{et al}. 1997). Consider an accelerator which prepares a
projectile and target in a particular state. The free particle Hamiltonian is
$H_{o}$ while the potential in the interaction region is $V$. The total
Hamiltonian modeling the interaction of the particle with the target is,
therefore, $H$ = $H_{o}$ + $V$.

An important step in their analysis of scattering experiment is the invocation
of the \textit{preparation/registration} arrow of time (Bohm \textit{et al}.
1994). The key intuition behind this arrow is that no observable properties
can be measured in a state until the state has been prepared. According to
Bohm it makes no sense to speak of a measurement of an observable such as the
scattering angle until there is a state prepared by the accelerator. The time
$t=0$ marks the moment in time at which the state preparation is completed and
the registration of detector counts can begin (any detector counts before this
time must be discarded as noise). One of the consequences of the
preparation/registration arrow is that some mathematical operations definable
in HS are nonsensical. For example one can calculate nonzero expectation
values for an observable for $t<0$, meaning that an observable has a nonzero
expectation value before the state has been prepared.

Following Ludwig (1983; 1985; Bohm et al. 1997), an in-state of a particular
quantum system (conceived of as an ensemble of individual systems such as each
elementary particle) is prepared by a preparation apparatus (a macrophysical
system). The detector (considered to be classical) registers the
post-interaction particles, also called out-states. In-states are taken to be
elements $\phi\in\Phi_{-}$ and observables are taken to be elements $\psi
\in\Phi_{+}$. (Decaying states, such as the Dirac, Lippman, Schwinger kets and
Gamow vectors, are elements of $\Phi_{+}^{\times}$).

The need to distinguish between states and observables implies the need for
two RHS's, one for the states and one for the observables. The RHS $\Phi
_{-}\subset\mathcal{H}\subset\Phi_{-}^{\times}$ is physically interpreted as
the space of states while the RHS $\Phi_{+}\subset\mathcal{H}\subset\Phi
_{+}^{\times}$ is physically interpreted as the space of observables. The
justification for these interpretations is as follows. The
preparation/registration arrow implies the mathematical conditions
$\int\langle E|\psi(t)\rangle dE=0$ for all $t<0$ and $\int\langle
E|\phi(t)\rangle dE=0$ for all $t>0$. The requirements of analytic
continuation leads naturally to a set of mathematical spaces fulfilling these
conditions: $\Phi_{-}$ is the Hardy space of the lower complex energy
half-plane intersected with the Schwartz class functions and $\Phi_{+}$ is the
Hardy space of the upper complex energy half-plane intersected with the
Schwartz class functions (e.g. Bohm et al. 1997).

For the space of states $\Phi_{-}$, we seek a continuous evolution operator
$U_{-}^{o}:\Phi_{-}\rightarrow\Phi_{-}$. $U$ restricted to $\Phi_{-}$ fulfils
this condition (i.e., it is continuous in $\tau_{\Phi_{-}}$), but only for
$t\leq0$. $U_{-}^{o}$ carries states into the forward direction of time.
Whereas $U$ forms a unitary group on HS, its restriction to the domain
$\Phi_{-}$ is a semigroup for times $t\leq0$. Since $U_{-}^{o}$ is $\tau
_{\Phi_{-}}$-continuous for times $t\leq0$, the extension of $U$ to $\Phi
_{-}^{\times}$ exists as a semigroup for $t\leq0$.

Similarly for the space of observables $\Phi_{+}$, we seek a continuous
evolution operator $U_{+}^{o}:\Phi_{+}\rightarrow\Phi_{+}$. $U$ restricted to
$\Phi_{+}$ fulfils this condition (i.e., it is continuous in $\tau_{\Phi_{+}}%
$) only for $t\geq0$ and its temporal direction carries observables into the
forward direction of time. Whereas $U$ is unitary on HS, its restriction to
the domain $\Phi_{+}$ is a semigroup for times $t\geq0$. Since $U_{+}^{o}$ is
$\tau_{\Phi_{+}}$-continuous for times $t\geq0$, the extension of $U$ to
$\Phi_{+}^{\times}$ exists as a semigroup for $t\geq0$.

$U$ extended to $\Phi_{-}^{\times}$ and $U$ extended to $\Phi_{+}^{\times}$
form two semigroups for which replacement of $t$ with $-t$ is not defined.
These semigroups fall out of the analysis quite naturally in the RHS framework
providing a rigorous description of irreversible behavior in a scattering
experiment (Bohm \textit{et al}. 1997).\footnote{In general $\Phi_{-}\neq
\Phi_{+}$, but $\Phi_{-}\cap\Phi_{+}$ $\neq\{0\}$, so the semi-groups derived
in this framework cannot be considered as leading to two disjoint families of eigenfunctions.}

As Bohm and Gadella (1989) demonstrate, some elements of the generalized
eigenstates in $\Phi_{-}^{\times}$ and $\Phi_{+}^{\times}$ correspond to
exponentially growing and decaying states respectively. The semigroups
governing these states are
\begin{subequations}
\begin{align}
\langle\phi|U^{\times}|Z_{R}^{\ast}\rangle &  =e^{-iE_{R}t}e^{\frac{\Gamma}%
{2}t}\langle\phi|Z_{R}^{\ast}\rangle\text{ }t\leq0\\
\langle\psi|U^{\times}|Z_{R}\rangle &  =e^{-iE_{R}t}e^{-\frac{\Gamma}{2}%
t}\langle\psi|Z_{R}\rangle\text{ }t\geq0\text{,}%
\end{align}
where states $\phi\in\Phi_{-}$, observables $\psi\in\Phi_{+}$, $E_{R}$
represents the total resonance energy, $\Gamma$ represents the resonance
width, $Z_{R}$ represents the pole at $E_{R}-i\frac{\Gamma}{2}$, $Z_{R}^{\ast
}$ represents the pole at $E_{R}+i\frac{\Gamma}{2}$, $|Z_{R}^{\ast}\rangle
\in\Phi_{-}^{\times}$ represents the growing Gamow vector and $|Z_{R}%
\rangle\in\Phi_{+}^{\times}$ represents the decaying Gamow vector. The $t<0$
semigroup is identified as future-directed along with $|Z_{R}^{\ast}\rangle$
as forming/growing states. The $t>0$ semigroup is identified as
future-directed along with $|Z_{R}\rangle$ as decaying states.

The preparation/registration arrow plays a crucial role in these
identifications, since it serves to specify the temporal direction of the
semigroups. The space of functions plus the semigroup property alone are
insufficient to determine the temporal direction of the semigroups. One can
object that relying on notions of preparation and registration are operational
or interventionist. Such an objection points to the good news/bad news nature
of Bohm and colleagues' work. The good news is that, given the highly
constrained context of the laboratory, operational procedures for preparations
and registrations can be spelled out precisely. Such an approach seems
justifiable for the scattering experiments of interest to Bohm. The bad news
is that the approach does not generalize straightforwardly to contexts outside
the laboratory.

\section{Semigroups in the Brussels-Austin Approach}

Prigogine and co-workers have also analyzed scattering and decay experiments
in their recent work. In their discussion of the Friedrichs model for
scattering and resonance phenomena, Antoniou and Prigogine apply the RHS
framework and show that the Hardy class functions form a natural function
space for the analysis of quantum scattering and decay phenomena (Antoniou and
Prigogine 1993). Unlike in Bohm's approach, however, they do not make Ludwig's
distinction between states and observables. Furthermore Antoniou and Prigogine
adopt the following time-ordering: excitations are interpreted as events
taking place before $t=0$ while de-excitations are to be interpreted as events
taking place after $t=0$. This arrow is a kind of generalization of the
preparation/registration arrow, but is based on observations rather than interventions.

The Brussels-Austin Group discusses two semigroups of evolution operators
acting on states in $\Phi^{\times}$. They split the test function space into
two spaces $\Phi_{-}$ and $\Phi_{+}$ based on their time-ordering rule. Upon
reaching the point where choices have to be made regarding interpreting the
directions of integration around the poles in the upper and lower complex
half-planes for the Hardy class functions, they make the following choices:
excitations (e.g. transitions from the continuum to the eigenstate in the
Friedrichs model, or formation of unstable states) are considered as
past-oriented and are associated with contours in the upper half-plane, while
de-excitations (e.g. mode-mode transitions in the Friedrichs model, or decay
of ustable states) are considered as future-oriented and are associated with
contours in the lower half-plane.

The eigenvectors of decaying states are associated with a discrete pole in the
continuum and are represented by elements in the dual spaces $\Phi_{-}%
^{\times}$ and $\Phi_{+}^{\times}$ (Antoniou and Prigogine 1993). By the same
continuity requirements as in Bohm's approach, the evolution operators split
into two time domains yielding
\end{subequations}
\begin{subequations}
\begin{align}
\langle\phi_{+}|U^{\times}|Z_{R}^{\ast}\rangle &  =e^{iE_{R}t}e^{\frac{\Gamma
}{2}t}\langle\phi_{+}|Z_{R}^{\ast}\rangle\text{ }t<0\\
\langle\phi_{-}|U^{\times}|Z_{R}\rangle &  =e^{-iE_{R}t}e^{-\frac{\Gamma}{2}%
t}\langle\phi_{-}|Z_{R}\rangle\text{ }t>0\text{,}%
\end{align}
where $\phi_{+}\in\Phi_{+}$ and $\phi_{-}\in\Phi_{-}$. Note that the roles of
the upper and lower Hardy class function spaces is reversed with respect to
Bohm's approach. The Brussels-Austin Group identifies the $t<0$ semigroup as
evolving states into the past along with $|Z_{R}^{\ast}\rangle$ as decaying
states, while the $t>0$ semigroup evolves states into the future along with
$|Z_{R}\rangle$ as decaying states.

As noted above, the space of functions plus the semigroup property alone are
insufficient to determine the temporal direction of the semigroups. The
Brussels-Austin Group uses consistency with both empirical observations, as
well as the ability of systems to communicate with each other, in order to
determine the directions of the semigroups (Antoniou and Prigogine 1993;
Antoniou, private communication).\footnote{The approach for transient
scattering can be extended to the case where the interactions are continuous
and persistent, yielding similar results (Petrosky and Prigogine 1997b).}

\section{Relating the Two Approaches}

There are two immediate observations when comparing the work of Bohm and
Brussels-Austin Groups: 1) The time directions identified for the $t<0$
semigroups differ between the two research groups. 2) The roles of the Hardy
class spaces are reversed. Both differences can be traced to the temporal
arrows and contexts invoked in the two approaches.

Bohm envisions the case of a scattering experiment, where the
preparation/registration arrow is built into the experimental arrangement by
the very nature of the interventions required. In its most general form this
arrow expresses the idea that observable properties do not exist apart from
some physical state, i.e., observable properties logically presuppose states.
However, the laboratory context accounts for the additional criteria allowing
Bohm to identify the two semigroups as both being future-directed. Hence,
there is a clearly motivated arrow of time, albeit in a limited context.

Antoniou and Prigogine envision a more general situation where excitations and
de-excitations of states occur in the absence of laboratory-type
interventions. Invoking such an arrow of time along with other conditions
(e.g. the ability to communicate) leads to the assignment of temporal
directions in (10). The $t<0$ semigroup is ignored as we never observe it,
leading to a consistent description of irreversible processes. Nevertheless,
the generality of the excitation/de-excitation arrow gives us no physically
rigorous argument for the temporal arrow because the arrow is supplemented
with conditions consistent with our experience.

One might be led to think that Antoniou's and Prigogine's not distinguishing
between states and observables leads to the differences between the two groups
since $\Phi_{-}\subset\Phi_{+}^{\times}$. However this is not the case,
because the distinction between states and observables is dependent upon the
preparation/registration arrow. Although preparations are particular kinds of
excitations and registrations are associated with particular kinds of
de-excitations, the Brussels-Austin focus on states leads naturally to a
different splitting of the RHSs based on their more general arrow.

\section{Interpreting the RHS Formalism}

Many advocates of RHS are reluctant to give a realistic interpretation to the
elements of the mathematical framework. Bohm and the Brussels-Austin Group are
likewise cautious in this regard, but indicate that they have some realist
leanings regarding these elements. For example both groups consider the
elements of $\Phi$ to represent possible physical states or observables of the
system (Bohm 1967; Antoniou and Prigogine 1993).\footnote{Indeed although both
$\Phi$ and $\mathcal{H}$ are mathematical idealizations, Bohm takes $\Phi$ to
be ``closer'' to reality than $\mathcal{H}$ (1978, 21).}

There are tensions, nevertheless, between what is considered realistic versus
what is considered merely useful for computational purposes. For example Bohm,
following Ludwig (1983; 1985), takes the preparation and registration
apparatuses to be classical devices. Preparation apparatuses prepare the
states $\phi$ while registration apparatuses detect or register the
observables (or the values thereof) $\psi$. These observables are considered
to be the ``real'' physical entities (Bohm \textit{et al}. 1997, 496-7). Bohm,
however, often asserts that microsystems--the ``agents by which the
preparation apparatus acts on the registration apparatus''--are imaginary
(e.g. Bohm \textit{et al}. 1994, 443) or that there is no need to assume they
exist (Bohm \textit{et al}. 1997, 496). The ``imagined entities connected with
microphysical systems are not restricted to $\Phi$; indeed their energy
distributions do not have to be well behaved functions of the energy'' (Bohm
\textit{et al}. 1997, 497). Therefore, Bohm concludes that, ``for the
hypothetical entities connected with microphysical systems, like Dirac's
`scattering states' $|p\rangle$ or Gamow's `decaying states' $|E-i\Gamma
/2\rangle$, the RHS formulation has a much larger choice and can describe them
by elements of $\Phi^{\times}$'' (Bohm \textit{et al}. 1997, 497; see also
Bohm \textit{et al}. 1994).

On the other hand, Bohm seems inclined at some points to attribute reality to
these entities: ``...there may be a larger class of `microphysical states' (in
addition to the Dirac Kets, Gamow vectors, virtual state vectors...and
others...) which still await their physical interpretation'' (Bohm \textit{et
al}. 1994, 446).\footnote{Bohm's point here is that there may be many other
elements in the space $\Phi^{\times}$ that have a physical correspondence or
interpretation (i.e. there may be new, as yet undescribed, physics), a point
on which he was silent in earlier writings on RHS quantum mechanics.} Or
again, ``Though there is an abundance of resonance states in nature which are
described by first-order Gamow vectors and which evolve according to the
exponential law, there is no direct experimental evidence for microphysical
objects associated with $N$th order poles, $N>1$, of the S-matrix which are
described by higher-order Gamow vectors'' (Bohm \textit{et al}. 1997, 529).
His comparison of the evidence for objects described by first order Gamow
vectors with the lack of evidence for objects described by higher-order Gamow
vectors appears to assume a realistic construal of such entities. Indeed, he
writes that the physical meaning of these higher-order Gamow vectors is
questionable in contrast with those of first-order (Bohm \textit{et al}. 1997,
532). Bohm then continues, ``ordinary Gamow states have been identified in
abundance, e.g., through their Breit-Wigner profile in scattering experiments,
or the exponential decay law'' (Bohm \textit{et al}. 1997, 532).

Although rarely making interpretive comments, the Brussels-Austin Group also
give indications of realist leanings. For example they assert that the
eigenvalues of observables in $\Phi^{\times}$ ``influence'' the evolution and
produces decay (Antoniou and Prigogine 1993, 454) and suggest that resonances
should be associated with physical observables in unstable systems (Antoniou
and Melnikov 1998). As well, they seek to reify distributions (elements of
$\Phi^{\times}$) as the fundamental ontological elements of descriptions in
both classical and quantum unstable systems (Petrosky and Prigogine 1997a;
1997b; Bishop 2004).

\section{Toward a Realistic Interpretation of RHS QM}

The RHS formalism has proven useful for illuminating our understanding of
particular irreversible processes found in a variety of unstable systems (e.g.
approach to equilibrium, decay, scattering). Yet the prospects for answering
our general questions about irreversibility and the origin of various arrows
of time are not clarified as yet. The preparation/registration arrow proposed
by Bohm clarifies the nature of irreversibility in scattering experiments, but
is much too limited for application to more general settings. For example when
the restriction to laboratory interventions is dropped--as in the
excitation/de-excitation arrow proposed by the Brussels-Austin Group--the
arrow can no longer uniquely determine the direction of the evolution
semigroups governing physical processes. We must still choose the temporal
directions of the semigroups based on additional criteria such as
observational experience or consistency.

Compared to the standard HS framework, the RHS framework provides a
significant advantage in the description of irreversible processes in that
semigroup evolutions arise naturally in the latter. Obviously more than the
presence of semigroups is needed, however, in order to explain the arrow of
time in quantum mechanics. One suggestion that could contribute to a more
complete understanding is to develop a robust realist interpretation of the
elements of the RHS formalism. A crucial reason why a realist interpretation
may prove important to further clarifying irreversibility and the quantum
arrow of time is provided by one of the core intuitions of the Bohm and
Brussels-Austin approaches: namely, that irreversibility is rooted in the
dynamics of physical systems. If that is the case, then the elements of the
RHS formalism have to be mapped onto elements of physical systems. So the
tensions discussed above need to be clarified and the realist suggestions
filled out in order to better elucidate the dynamical mechanisms at work in
irreversible processes.

A realist interpretation of the elements of the RHS formalism cannot be
carried out generically (see Melsheimer 1974a and 1974b for some indications
why). It requires concrete realizations of the dual pair $\{\Phi,\Phi^{\times
}\}$ which are tied to the algebra of observables of the systems in question.
Once such an interpretation for a given dual pair is in place, the power of
the RHS framework for clarifying and illuminating the dynamical processes
responsible for irreversible behavior of these systems should be greatly
enhanced. Some evidence for this can be seen in (Bohm \textit{et al}. 1997;
Petrosky and Prigogine 1997a and 1997b; Bishop 2004).

For example in RHS quantum mechanics the decay of scattering states is
associated with a Gamow vector with eigenvalue $\lambda=E_{R}-i\frac{\Gamma
}{2}$, a mathematical element not well defined on HS. The Gamow vector
involves physical quantities such as the resonance energy and the full width
at half maximum (note that $\Gamma=0$ corresponds to the rest energy for the
composite particle). Furthermore, under the Bohm approach, the condition
$\langle E|\psi(t)\rangle=0$ for all $t<0$ refers to the energy distribution
of the detected state while $\langle E|\phi(t)\rangle=0$ for all $t>0$ refers
to the incident beam resolution. All of these quantities are physically
measurable; however the concepts involved (energy, momentum, time, etc.) are
not exhausted simply by associating them with preparation or measurement
procedures. Hence, in the RHS framework, one can then make a direct
correspondence between mathematical elements, on the one hand, and their
physical counterparts and causal efficacy, on the other hand, in a way that
goes beyond operational procedures.

Other viewpoints focus on initial conditions as the explanation for
irreversibility and arrows of time. However, realistic initial conditions
involved in explaining irreversibility and time arrows often cannot be
formulated in HS. For example in the case of scattering, the standard initial
condition in HS is that states are not interacting with the scattering center
at $t\rightarrow-\infty$. Of course this initial condition is unrealistic as
the particles crucial to the experiment have not been created or properly
prepared until some finite time before the interaction. Yet HS cannot
accommodate more physically realistic initial conditions for scattering
processes (Bohm \textit{et al}. 1997). The RHS framework can accommodate
realistic initial conditions in a natural way; so a realistic interpretation
of the elements of RHS could also play a fruitful role for the initial
condition route to explanations of irreversibility as well.

\section{Refernces}

Antoniou, I. and Melnikov, Yu. (1998), ``Quantum Scattering of Resonances:
Poles of a Continued S-Matrix and Poles of an Extended Resolvent,'' in Bohm et
al. (1998).

Antoniou, I. and Prigogine, I.. (1993), ``Intrinsic Irreversibility and
Integrability of Dynamics,'' \textit{Physica A} 192: 443-64.

Atmanspacher, H. Bishop, R. and Amann, A. (200), ``Extrinsic and Intrinsic
Irreversibility in Probabilistic Dynamical Laws,'' in L. Accardi (ed.)
\textit{Quantum Probability and Related Topics}. Singapore: World Scientific,
pp. 50-70.

Bishop, R. (2004), ``Nonequilibrium Statistical Mechanics
Brussels-AustinStyle,'' \textit{Studies in History and Philosophy of Modern
Physics} 35:1-30.

Bohm, A. (1967), ``Rigged Hilbert Space and Mathematical Description of
Physical Systems,'' in W. Brittin, et al. (eds.) \textit{Lectures in
Theoretical Physics Vol IX A: Mathematical Methods of Theoretical Physics}.
New York: Gordon and Breach Science Publishers.

Bohm, A. (1978), \textit{The Rigged Hilbert Space and Quantum Mechanics}.
Berlin: Springer-Verlag.

\_\_\_\_\_\_\_ (1993), \textit{Quantum Mechanics: Foundations and
Applications}. Berlin: Springer-Verlag.

Bohm, A., Antoniou, I. and Kielanowski, P. (1994), ``The
preparation/registration Arrow of Time in Quantum Mechanics,'' \textit{Physics
Letters A} 189:442-8.

Bohm, A., Doebner H-D. and Kielanowski, P. (1998), \textit{Irreversibility and
Causality: Semigroups and Rigged Hilbert Spaces}. Berlin: Springer-Verlag.

Bohm, A. and Gadella, M. (1989), \textit{Dirac Kets, Gamow Vectors, and
Gel'fand Triplets, Lecture Notes in Physics, vol. 348}. Berlin: Springer-Verlag.

Bohm, A. Gadella, M. and Wickramasekara, S. (1999), ``Some Little Things about
Rigged Hilbert Spaces and Quantum Mechanics and All That,'' in I. Antoniou and
G. Lumer (eds.) \textit{Generalized Functions, Operator Theory, and Dynamical
Systems}. Boca Raton, FL: Chapman \& Hall/CRC, 202-250.

Bohm, A., Maxson, S. Loewe, M. and Gadella, M. (1997), ``Quantum Mechanical
Irreversibility,'' \textit{Physica A} 236: 485-549.

Dirac, P. (1981/1930), \textit{The Principles of Quantum Mechanics Fourth
Edition}. Oxford: The Clarendon Press.

Gadella, M. (1983), ``A Rigged Hilbert Space of Hardy-Class Functions:
Applications to Resonances,'' \textit{Journal of Mathematical Physics} 24: 1462-9.

Gel'fand, I. and Shilov, G. (1967), \textit{Generalized Functions Volume 3:
Theory of Differential Equations}, M. Mayer tr. New York: Academic Press.

Gel'fand, I. and Vilenkin, N. (1964), \textit{Generalized Functions Volume 4:
Applications of Harmonic Analysis}, A. Feinstein tr. New York: Academic Press.

Ludwig, G. (1983), \textit{Foundations of Quantum Mechanics, Vol. I}. Berlin: Springer-Verlag.

\_\_\_\_\_\_\_\_ (1985), \textit{Foundations of Quantum Mechanics, Vol. II.}
Berlin: Springer-Verlag.

Melsheimer, O. (1974a), ``Rigged Hilbert Space Formalism as an Extended
Mathematical formalism for Quantum Systems. I. General Theory,''
\textit{Journal of Mathematical Physics} 15: 902-16.

\_\_\_\_\_\_\_\_\_\_\_\_ (1974b), ``Rigged Hilbert Space Formalism as an
Extended Mathematical formalism for Quantum Systems. II. Transformation Theory
in Nonrelativistic Quantum Mechanics,'' \textit{Journal of Mathematical
Physics} 15: 917-26.

Nagel, B. (1989), ''Introduction to Rigged Hilbert Spaces,'' in E. Br\"{a}ndas
and N. Elander (eds.) \textit{Resonances: The Unifying Route Towards the
Formulation of Dynamical Processes: Foundations and Applications in Nuclear,
Atomic, and Molecular Physics}. Berlin: Springer-Verlag..

Nelson, E. (1959),. ``Analytic Vectors,'' \textit{Annals of Mathematics} 70: 572-615.

Petrosky, T. and Prigogine, I (1997a), ``The Liouville Space Extension of
Quantum Mechanics,'' in I. Prigogine and S. Rice (eds.) \textit{Advances in
Chemical Physics Volume XCIX}. New York: John Wiley \& Sons.

Petrosky, T. And Prigogine, I. (1997b), ``The Extension of Classical Dynamics
for Unstable Hamiltonian Systems,'' \textit{Computers \& Mathematics with
Applications} 34: 1-44.

Treves, F. (1967), \textit{Topological Vector Spaces, Distributions and
Kernels}. New York: Academic Press.

von Neumann, J. (1955/1932), \textit{Mathematical Foundations of Quantum
Mechanics}. Princeton: Princeton University Press.

Wilkinson, S.,Bharucha, F., Fischer, M., Morrow, P., Niu, Q., Sundaram, B. and
Raizen, M. (1997), ``Observation of Non-Exponential Decay in Quantum
Tunneling,'' \textit{Nature} 387: 575-77.
\end{subequations}
\end{document}